\documentstyle[prl,aps,multicol,epsf]{revtex}
\begin{document}
\draft
\title{Hubbard model versus $t$-$J$ model: The one-particle spectrum}
\author{Henk Eskes}
\address{Institute Lorentz, Leiden University, P.O. Box 9506, 
         2300 RA Leiden, The Netherlands}
\author{Robert Eder}
\address{Laboratory of Applied and Solid-State Physics, 
         Materials Science Centre,\\ 
         University of Groningen, Nijenborgh 4,
         9747 AG Groningen, The Netherlands}
\date{\today}
\maketitle

\begin{abstract}
The origin of the apparent discrepancies between the one-particle 
spectra of the Hubbard and \mbox{$t$-$J$} models is revealed:
Wavefunction corrections, in addition to the three-site terms, 
should supplement the bare \mbox{$t$-$J$} model. In this way a quantitative 
agreement between the two models is obtained, even for the 
intermediate-$U$ values appropriate for the high-$T_{c}$ cuprate 
superconductors.  Numerical results for clusters of up to 20 sites 
are presented.  The momentum dependence of the observed intensities 
in the photoemission spectra of Sr$_2$CuO$_2$Cl$_2$ are well described 
by this complete strong-coupling approach.
\end{abstract}


\begin{multicols}{2}

The past ten years the interest in Hamiltonians modeling correlated 
electrons has greatly increased.  This is largely due to 
the mysterious normal-state properties of the high-$T_{c}$ cuprate 
superconductors and the realization that local correlations on the Cu 
sites are large in these compounds.  The two most intensively studied 
Hamiltonians are the Hubbard and \mbox{$t$-$J$} models. 
Although at first sight 
such models may seem oversimplified, it has been argued that the 
two dimensional square-lattice Hubbard model captures the essential 
dynamics of the cuprates in a window of a few $eV$ around the chemical 
potential\cite{And87,Zha88,Hyb90,Fei96}. The materials belong to the 
intermediate-$U$ regime, $U/t \approx 10$, $t \approx 0.3-0.4$ 
eV\cite{Hyb90,Fei96}. Further support for a single-band Hubbard approach
comes from measured one-particle and optical spectra of the cuprates. 
These show a dramatic reshuffling of intensity as a function of 
doping on the scale of a few eV \cite{Che91,Uch91}. This characteristic 
behavior is a fingerprint of strong correlations and is described well
by the Hubbard model\cite{Che91,Dag94,Esk94}.

Although the \mbox{$t$-$J$} model is frequently referred to
as being the large-$U$ 
limit of the single-band Hubbard model, the (one-particle) spectra of 
the two models differ significantly in the intermediate-$U$ 
regime of interest. (For a recent review on numerical results, see 
Ref.~\onlinecite{Dag94})  Probably because of this, the two models 
are often treated as being ``distinct''.  One obvious difference 
concerns the momentum occupation number $n_{\bf k}$ at half filling, 
equal to the one-electron removal sum rule.  For the \mbox{$t$-$J$} model 
this quantity is trivially equal to $1/2$, while for the 
intermediate-$U$ Hubbard model it is a strongly varying function of 
${\bf k}$.  Even though in the infinite-$U$ limit $n_{\bf k}$
tends to $1/2$, as in the \mbox{$t$-$J$} model, the convergence to this 
value by increasing $U$ is very slow.

In this Letter we show that these discrepancies can be attributed to
the neglect of certain terms in the large-$U$ 
perturbation theory.  Specifically, the weight distributions in the 
spectral functions show a high sensitivity to the first-order 
renormalization of the wave functions.
The complete strong-coupling approach is found to quantitatively reproduce 
the Hubbard spectra for $U \geq W$, where $W$ is the $U=0$ band width.
For these $U$ values the perturbation series in $t/U$ agrees well
with the full Hubbard Hamiltonian.

The derivation of the strong-coupling Hamiltonian by means 
of a canonical transformation, written as an expansion in $t/U$, is 
well known\cite{Esk94,Har67}. Starting point is the Hubbard model,
\begin{equation}
H = V + T =
   U \sum \limits _{i} n_{i,\uparrow} n_{i,\downarrow} 
 - t \sum\limits_{i,\delta,\sigma}
    a^{\dagger}_{i,\sigma} a_{i+\delta,\sigma}.             \label{hub}
\end{equation}
Here $n_{i,\uparrow}$ is the occupation number of the original Hubbard 
fermions $a_{i,\sigma}$, and ${\bf \delta}$ is a vector connecting 
nearest neighbors.  After the transformation the states in the lowest 
Hubbard sector (no doubly occupied sites) are described to order 
$t^2/U$ by the strong-coupling Hamiltonian,
\begin{eqnarray}
&&H_{sc} = - t \sum\limits_{i,\delta,\sigma}
          c^{\dagger}_{i,\sigma} c_{i+\delta,\sigma} 
  -  \frac{J}{2} \sum\limits_{i,\delta} (
       \tilde{\bf S}_{i} \cdot \tilde{\bf S}_{i+\delta} -
       \frac{1}{4} \tilde{n}_{i} \tilde{n}_{i+\delta}   )      \nonumber \\
&&-  \frac{J}{4} \sum\limits_{i,\delta\neq\delta^{\prime},\sigma}
     \left(
         c^{\dagger}_{i+\delta,\sigma} \tilde{n}_{i,\bar{\sigma}}
         c_{i+\delta^{\prime},\sigma} 
      -  c^{\dagger}_{i+\delta,\bar{\sigma}}
         c^{\dagger}_{i,\sigma} c_{i,\bar{\sigma}} 
         c_{i+\delta^{\prime},\sigma}
     \right),                                                 \nonumber \\
                                                              \label{sc}
\end{eqnarray}
where $J = 4t^2/U$.  The motion of the above transformed 
$c_{i,\sigma}$ fermions will leave 
the number of doubly occupied sites unchanged, and 
$\tilde{\bf S}_{i}$ and $\tilde{n}_{i,\sigma}$ are the spin and 
occupation number of these new fermions.  Simply 
neglecting the `three-site term', i.e.\ the second line
of (\ref{sc}), one obtains the \mbox{$t$-$J$} model. 

Why the full strong-coupling model has received far less attention 
than the \mbox{$t$-$J$} model is unclear.  
It might be expected that the three-site term is very
important in the one-hole limit (and low doping region)
since, unlike the nearest-neighbor hopping,
a direct next-nearest-neighbor hopping process does not 
frustrate the spin system. However, it was shown by a numerical 
study\cite{Szc90}, spin-polaron approaches\cite{Tru88,Ede90} 
and a linear-spin-wave self-consistent Born (LSW-SCB)
calculation\cite{Bal95} that the main effect of this term on the 
quasiparticle dispersion is an increase of the bandwidth
of the dominant `quasiparticle peak'. 

By virtue of its construction via a canonical
transformation, the Hamiltonian (\ref{sc}) reproduces the
eigenvalue spectrum of the Hubbard model for large $U$. To obtain further
information, such as matrix elements or expectation values,
operators acting in the Hilbert space of the Hubbard model
have to be subjected to the same canonical transformation.
For the electron annihilation operator this
procedure leads to the expression\cite{Esk94,Har67},
\begin{eqnarray}
&P& a_{i,\sigma} P =  P   c_{i,\sigma}   P       \nonumber \\
& & + \frac{t}{U} \sum\limits_{\delta}
      P \left(
      \tilde{n}_{i,\bar{\sigma}} c_{i+\delta,\sigma}
   -  c^{\dagger}_{i,\bar{\sigma}} c_{i,\sigma} c_{i+\delta,\bar{\sigma}} 
      \right)                        P.                  \label{a}
\end{eqnarray}
$P$ is defined as the projector on states with no double occupancy of 
the transformed fermions 
(states with $\langle \tilde{V} \rangle = 0$).
The first-order correction terms (second line) describe the
dressing of the annihilation operator with `virtual' charge
fluctuations involving doubly-occupied sites.
While their prefactor $t/U$ seems to suggest that they
represent only a minor correction, their impact on the calculated
spectra is dramatic -- one reason being that the summation
over neighbors introduces the 
coordination number $z$ as an additional `hidden' prefactor. 
When comparing strong-coupling calculations directly with experiments 
it is therefore crucial to take these corrections 
into account.  For the optical conductivity such corrections 
were shown to lead to qualitatively different results\cite{Ste92}.
Another consequence is a
rapid reshuffling of weight between the upper (UHB) 
and lower-Hubbard band (LHB) as a function of hole doping, observed 
in both one-particle and optical experiments.
The first-order corrections to transformed operators 
such as Eq.~(\ref{a}) lead to modified expressions for sum rules 
that consistently describe the rapid intensity changes observed 
in experiments\cite{Che91,Uch91,Esk94,Har67}.

For the numerical calculations of the one-particle spectra we 
implemented the strong-coupling model Eq.~(\ref{sc}).  Electron removal 
is described by the operator Eq.~(\ref{a}).  This operator does not 
change double-occupancy and only the LHB part of the Hubbard spectrum 
is produced.  (There is, in fact, a similar independent procedure to 
calculate the UHB response as well.)  The $t/U$ term in Eq.~(\ref{a}) 
amounts to the removal of an electron from a neighbor of $i$.  
This leads to an extra phase factor $\exp(i{\bf k}\cdot{\bf \delta})$.  
Because of this the total weight is enhanced at $\Gamma$ and reduced
at $(\pi,\pi)$ and Eq.~(\ref{a}) leads to the correct 
behavior of $n({\bf k})$ (see Fig.~2 of Ref.~\onlinecite{Esk94}).  
Note that the implementation of the first-order 
correction is straightforward and does not increase memory or CPU time 
in 
\begin{figure}
\epsfxsize=\hsize
\vspace{0in}
\hspace{0ex}\epsffile{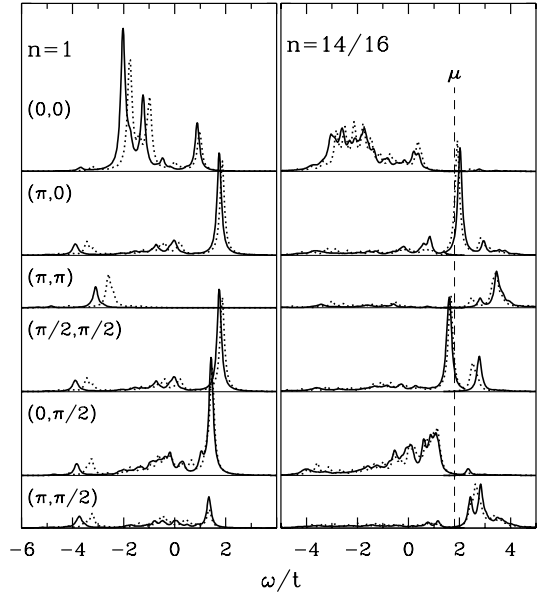}
\vspace{0ex}
\narrowtext
\caption[]{ The one-particle spectrum of the two-dimensional 4x4 
strong-coupling model (solid line) compared with the Hubbard model 
(dotted line), at half filling (left) and for two holes (right).  
$U/t=10$.  Only the LHB part of the Hubbard spectrum is shown.  Dashed 
line is the chemical potential in the two-hole case.  Hubbard data 
reproduced from Ref.~\protect\onlinecite{Leu92} }
\label{fig_16} 
\end{figure}
\noindent a Lanczos-like diagonalization calculation.  In a similar way 
such operator corrections can be taken into account in analytic approaches, 
for instance in the LSW-SCB calculation of the one-hole spectrum.
  
In Fig.~\ref{fig_16} we compare the two-dimensional 16-site 4x4 
strong-coupling model with the LHB electron removal and addition 
spectra of the Hubbard model, reproduced from Ref.~\onlinecite{Leu92}.  
Note that $U=10$ is comparable to the bandwidth $W=8$.  No 
shift or scaling has been applied to the spectra, showing that the value 
of the gap between the UHB (not shown, starting at energy $\approx 8t$) and 
LHB is accurately described. (While there is no UHB in
the strong coupling model, the magnitude of the energy gap
still can be inferred from the relation
$\Delta = U - 2\cdot E_{min}$, with $E_{min}\approx 2t$ the minimal
ionization energy of the model.)
The resulting gap of $\approx 6t$ is consistent with
the measured insulating gap of $\approx 2 eV$.

As can be seen, excitation energies and weights of the dominant
features in the spectra are
reproduced correctly by the strong-coupling approach.
The agreement is best for the physically most relevant
low excitation energies, but
the strong coupling approach also gives a complete 
description of the higher excitations, with energies not 
much smaller than $U$.  The total width is slightly overestimated by the 
second-order $t^2/U$ terms in the Hamiltonian, due especially to those 
peaks which obtain a large energy correction by the three-site terms, 
for instance at $(\pi,\pi)$. 
There is no significant deterioration of the agreement upon doping,
demonstrating that the discrepancies between the doping dependences
of low energy sum-rules for
$t-J$ and Hubbard originate from matrix element effects,
rather than an incompleteness of the Hilbert space of the
strong coupling model.  We explicitly checked that second-order 
corrections to the Fermi operators lead to only minor changes of the 
peak intensities.  For $U$ values below $10$ 
the agreement between Hubbard and its strong-coupling limit gradually
deteriorates, which we interpret as a breaking down of the 
perturbation approach.

Stimulated by the good agreement in the 4x4 system we now present 
large-U Hubbard 
spectra for clusters up to 20 sites.  Because the momenta of the 16 and 18 
site clusters nicely complement each other and are all on the
high-symmetry axis, we show a mix of these two clusters in Figs.~2 and 
3.  The curves on the left show a scan from $(0,0)$ to 
$(\pi,\pi)$.  The plots on the right go from $(0,0)$ to $(\pi,0)$ and 
from $(\pi,0)$ to $(\pi,\pi)$.  We shifted all $4\times 4$ spectra by 
the same small energy such that the lowest peak at ${\bf k}=(0,0)$ is 
at the same energy for both clusters.  
For the two-hole case 
we shifted the \mbox{$t$-$J$} data by an additional $0.3t$ to have the 
chemical potentials at roughly the same position. The ground-state 
momentum is $(0,0)$ in all cases (there is an extra degeneracy for 
16-sites, 2 holes). In both figures the results are compared with the 
``bare'' \mbox{$t$-$J$} model.

As mentioned before there are substantial differences between \mbox{$t$-$J$} 
and strong coupling.  At half-filling the low energy `quasiparticle peaks' 
are
at higher energies (implying a larger Hubbard gap) for the \linebreak[4]
\begin{figure}
\epsfxsize=\hsize
\vspace{-0.16in}
\hspace{0ex}\epsffile{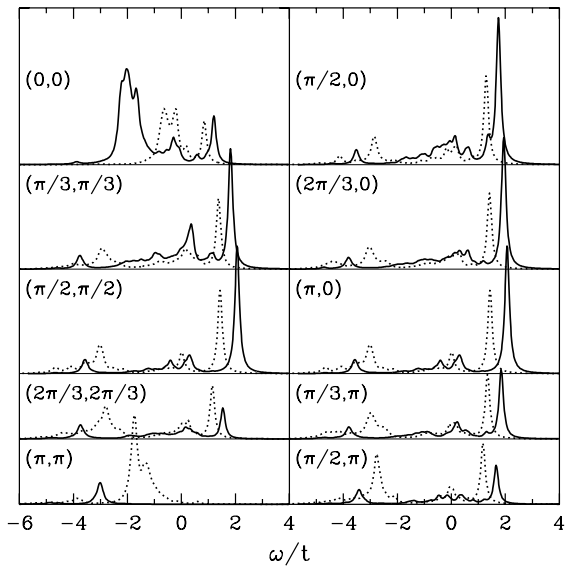}
\vspace{0ex}
\narrowtext
\caption[]{ The momentum dependence of the electron-removal spectrum of 
the strong-coupling Hubbard model (solid line) compared with the 
\mbox{$t$-$J$} model (dotted line).  The plot is a mix of data from the 16 
and 18 site clusters.  $U/t=10$.  $n=1$. 
The spectra for ${\bf k}=(\pi/2,\pi/2)$, $(\pi/2,0)$, $(\pi,0)$ and 
$(\pi/2,\pi)$ are calculated using the 2D 16-site cluster; the others, 
using the 18-site cluster.  }
\label{fig_1618} 
\end{figure}

\begin{figure}
\epsfxsize=\hsize
\vspace{0ex}
\hspace{0ex}\epsffile{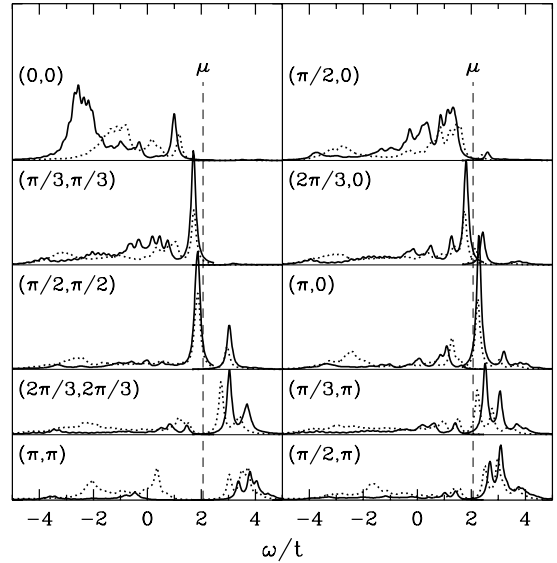}
\vspace{0ex}
\narrowtext
\caption[]{ Combined electron addition and removal spectra for the 
two-hole ground state.  See caption of Fig.~\protect\ref{fig_1618}.
}
\label{fig_2hole} 
\end{figure}
\noindent \mbox{$t$-$J$} model and their dispersion has a smaller
width.  Moreover, the momentum dependence of the weights is qualitatively
different.  The \mbox{$t$-$J$} spectrum at $(0,0)$, and
indeed through much of the Brillouin zone, is more 
compact than for large-$U$ Hubbard.  The strongest difference is at 
$(\pi,\pi)$.  Here strong coupling shows a little peak at high energy, 
while the \mbox{$t$-$J$} model has pronounced structures at a more 
intermediate energy. This reflects the small value of $n_{\bf k}$ 
for ${\bf k} = (\pi,\pi)$ in the Hubbard model, which is 
reproduced by the first-order corrections to the Fermi operator
Eq.~(\ref{a}). 

The spectra at half filling can be compared with recent angle-resolved 
photoemission data on insulating Sr$_2$CuO$_2$Cl$_2$\cite{Wel95}. 
A significant feature of these measurements is the strong momentum 
dependence of the weights of the spectra. 
The region in momentum space with the most intense
peaks is not centered around $(\pi/2,\pi/2)$, but displaced towards $(0,0)$.
Beyond $(\pi/2,\pi/2)$ the intensity drops very sharply and seems 
to vanish around $(\pi,\pi)$\cite{pes}.  As shown in Fig.~4 these features 
are well reproduced by the strong-coupling Hubbard model.
In contrast, the intensities in \mbox{$t$-$J$} are much more symmetric
around $(\pi/2,\pi/2)$ and change only little in that part of momentum
space. On average the peaks in strong-coupling Hubbard are more 
intense than those in the \mbox{$t$-$J$} model, at the expense of the
incoherent part of the spectrum.
Note that also in the \mbox{$t$-$J$} model there are differences 
in weight between ${\bf k}$ and ${\bf k} + (\pi,\pi)$.  This is due to
the coupling of the created hole to quantum-spin fluctuations
in the ground state of the Heisenberg antiferromagnet\cite{Ede90}. 
Experimentally the strong changes at the antiferromagnetic 
Brillouin-zone boundary look like  \linebreak[4]
\begin{figure}
\epsfxsize=\hsize
\vspace{-0.16in}
\hspace{0ex}\epsffile{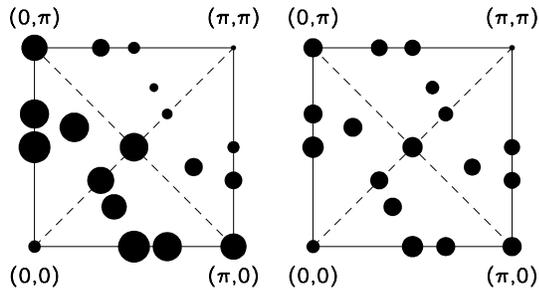}
\vspace{0ex}
\narrowtext
\caption[]{ Weights (proportional to the radius of the circles) of the 
lowest energy peak for the strong-coupling Hubbard model 
(left) and \mbox{$t$-$J$} model (right).  Data collected from the 16, 18 and 
20 site clusters.  $U/t=10$, $n=1$.  }
\label{fig_weight}  
\end{figure}
\noindent a kind of `pseudo Fermi surface': The
spectral weight observed drops significantly, but
an actual Fermi-level crossing does not take place.
We note that the large difference observed between $(\pi,0)$ and 
$(\pi/2,\pi/2)$ in 
the photoemission measurements actually is in disagreement 
with the near degeneracy of these momenta in both the standard Hubbard 
and \mbox{$t$-$J$} model.  As discussed by various authors this may be
explained by adding extra next-neighbor hopping terms to the 
Hamiltonian\cite{Bal95,Naz95}.

The increased quasiparticle weight and the strong drop of intensity when 
passing the half-filled Fermi surface are consistent with a continuous 
cross-over to a small-$U$ spin-density-wave (SDW) like shadow-band 
scenario\cite{Kam90}. Surely Fig.~\ref{fig_weight} shows that such a 
strong drop in weight is not necessarily a weak-coupling feature.

The interpretation of the doping dependence of the spectral function 
is a subtle issue\cite{Ste91,Pre95,Ede94,Wen96}, and will not be discussed 
here. For roughly 10\% holes (Fig.~\ref{fig_2hole}) there are intense 
features appearing in the electron addition spectrum for all 
momenta outside the magnetic zone. These features are more intense than 
for \mbox{$t$-$J$} and because of the sum rule the extra weight must come 
from the upper Hubbard band. This weight-transfer effect has been 
clearly observed in O-1s X-ray absorption\cite{Che91}. 

In conclusion, we have shown that the strong-coupling approach is 
a good alternative to Hubbard calculations even in the intermediate $U$ 
regime relevant for the high-$T_{c}$ superconductors. Only the 
{\it complete\/} strong-coupling model gives a faithful representation 
of the properties of the Hubbard model. The corrections 
to \mbox{$t$-$J$} lead to a smooth crossover to small $U$.  For comparisons 
with experiments such as angle-resolved photoemission, O-1s X-ray 
absorption (and optical conductivity) it is important to use the 
renormalized wavefunctions (rotated operators), as these can lead to even 
qualitatively different results.

We acknowledge stimulating discussions with George A. Sawatzky, Jan Zaanen 
and Andrzej M. Ole\'{s}.  HE is supported by the Stichting voor Fundamenteel 
Onderzoek der Materie (FOM), which is financially supported by the 
Nederlandse Organisatie voor Wetenschappelijk Onderzoek (NWO).
RE is supported by a postdoctoral fellowship granted by the Commission of 
the European Communities.

\end{multicols}


\begin{references}
\bibitem{pes} It is important to realize that other effects  
              can influence the momentum dependence of intensities as well, 
              such as the changing angle of the escaping electron 
              and the actual underlying Cu-3d, O-2p orbital structure.
\end{references}
\end{document}